\documentclass{an}
\usepackage{graphicx}
\usepackage{times}
\usepackage{fancyhdr}
\begin{document}

\title{New radio halos and relics in clusters of galaxies}

\author{G. Giovannini\inst{1,2}, L. Feretti\inst{1}, F. Govoni\inst{1,3},
M. Murgia\inst{1,3}, \and R. Pizzo\inst{1,4}} 

\institute{
INAF, Istituto di Radioastronomia, via Gobetti 101, 40129 Bologna, Italy
\and 
Dipartimento di Astronomia, Universita' di Bologna, via Ranzani 1, 40127 
Bologna, Italy
\and 
INAF, Osservatorio Astronomico di Cagliari, Loc. Poggio dei Pini, Strada 54,
09012 Capoterra (CA), Italy
\and
Kapteyn Astronomical Institute, Postbus 800,
9700 AV Groningen, Netherlands
}

\date{Received $<$date$>$; 
accepted $<$date$>$;
published online $<$date$>$}

\abstract{We present here new images of relics and halo sources in rich cluster
of galaxies and the correlation between the halo radio surface 
brightness versus the cluster bolometric X-ray luminosity.
}

\correspondence{ggiovann@ira.inaf.it}

\maketitle

\section{New radio halos and relics}
The knowledge of magnetic fields in clusters of galaxies is
important to confirm the existence of large-scale
cosmological magnetic fields and to study their properties.
It is therefore crucial to obtain detailed observations of radio
halos and relics, since these extended sources can give observational 
constraints to the cluster magnetic fields strength and distribution.
We present here new data for three clusters of galaxies:
A209, A548b, and A1758. From the diffuse radio sources in these clusters,
equipartition magnetic fields of the order of 0.5 $\times$ 10$^{-6}$ Gauss
are derived \footnote{we use H$_0$ = 70 km sec$^{-1}$ Mpc$^{-1}$; 
$\Omega_m$=0.3,and $\Omega_\Lambda$=0.7}.

{\bf Abell 209}.
 This cluster at z = 0.206 is dominated by a central cD galaxy located near
the peak
of the cluster mass distribution.
X-ray data and optical density distribution
show an irregular morphology with significant
substructures suggesting the presence of a cluster merger (Mercurio et al. 
2003). The VLA radio image
after the subtraction of discrete sources (Figure 1)
shows a giant diffuse radio halo  at the cluster center,
with a major axis size of $\sim$ 6' ($\sim$ 1.2 Mpc). 

\begin{figure}
{\includegraphics[angle=0,width=7.7cm,bb=36 110 576 660]{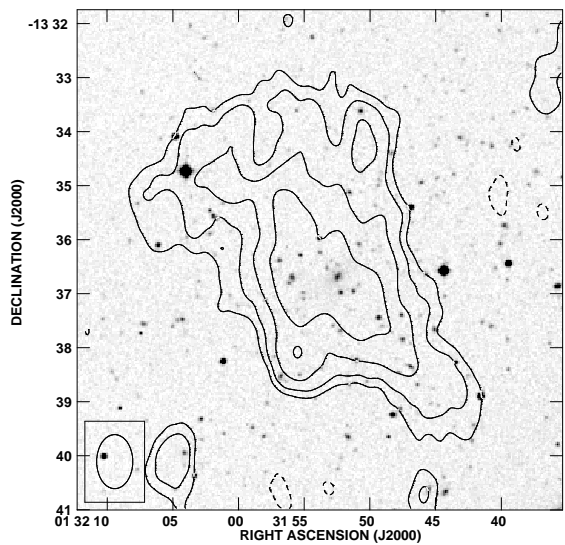}}
\caption{VLA radio image (contours) of the giant halo in
A209 overimposed to the optical map
from the Digital Sky Survey (gray scale). 
The HPBW is 40''$\times$60''; contours are: -0.2,0.2,0.3,0.5,0.8 mJy/beam. 
At the cluster distance, 1' $\sim$ 200 kpc.}
\label{a209}
\end{figure}

{\bf Abell 548b}.
The cluster A548, at an average redshift z = 0.04, shows a rather complex
structure with at least three main sub-clusters. The combination of X-ray and
optical data indicates that A548 is a cluster in a collapsing phase and
therefore not yet dynamically relaxed.
Feretti et al. (2006) report the detection of two diffuse relic radio
sources in A548b, confirming the association
between relics and mergers.
These sources are located on the same side of the
cluster's X-ray
center at projected distances $\sim$ 430 and 500 kpc, and show a similar
shape, flux density, extent, polarization, and spectral index. Another diffuse
source, probably a third relic, is detected close in projection to the
cluster center (Figure 2).

\begin{figure}
{\includegraphics[width=7.7cm,bb=0 10 432 470]{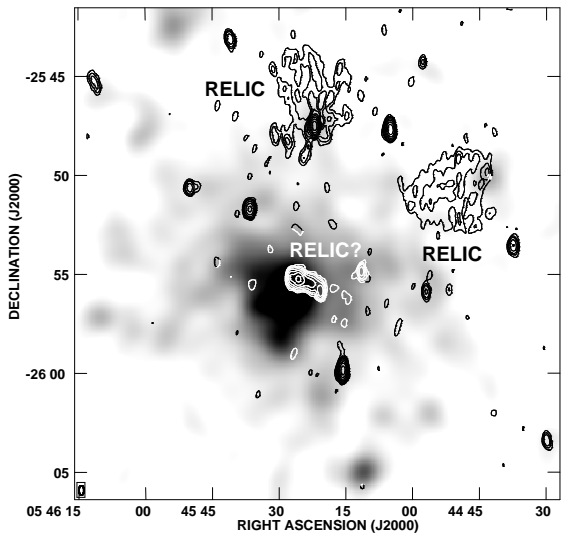}}
\caption{Overlay of the VLA radio image (contours) onto the cluster
X-ray image from ROSAT PSPC (gray scale). The radio HPBW is
15''$\times$30''. Contours are: 0.3,0.5,1,2,4,8,16,32,64 mJy/beam.
At the cluster distance, 1' $\sim$ 50 kpc.}
\label{a548}
\end{figure}

{\bf Abell 1758}. This cluster at z = 0.279 was studied in detail in the 
X-ray band by David \&
Kempner (2004).
They showed that A1758 consists of two
hot X-ray luminous clusters: A1758N and A1758S. The northern
cluster is in the later
stages of a merger of two 7 keV clusters, while A1758S is in the earlier
stages of a merger of two 5 keV clusters.

We have detected a radio halo in A1758N. 
In Figure 3 we show a superposition of the halo source
(discrete sources have been subracted) and the XMM cluster image.
The radio image shows a diffuse emission (central halo) permeating the
A1758N region where
the two subclusters are merging, $\sim$ 0.8 Mpc in size,
and two brighter peripheral structures on the opposite side with respect to the
cluster center, which resemble relic radio sources.
The detection of a radio diffuse emission in A1758N and not in A1758S is
in agreement with the hotter temperature (7keV) of the N subclusters with
respect to S subclusters (5 keV) and with the David \& Kempner result that
A1758N is in a late stage of merger, while A1758S is in an early merger stage.

\begin{figure}
{\includegraphics[angle=0,width=8.5cm, bb=36 130 576 658]{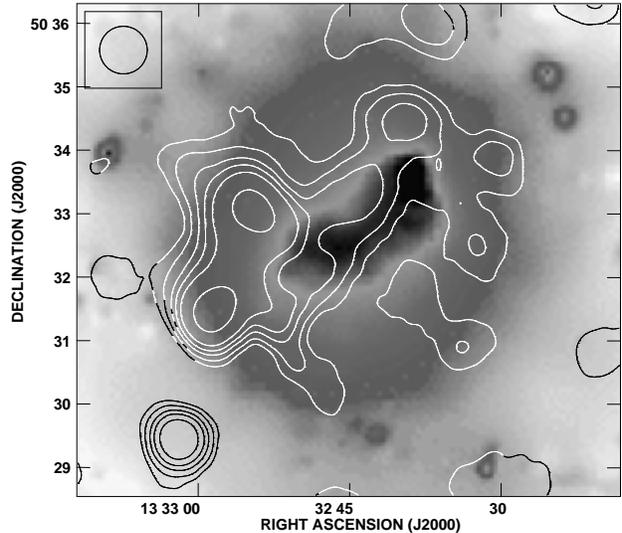}}
\caption{VLA radio image (contours) of the extended radio structure in 
A1758N overimposed to the XMM X-ray image (gray scale). 
The HPBW is 45''; contours are: 0.15,0.3,0.5,0.7,1,2,3,5,7 mJy/beam.
At the cluster distance, 1' $\sim$ 250 kpc}
\label{a1758}
\end{figure}

\section{Correlations}

To understand the origin, evolution, and physical properties of radio halos
and relics it is important to correlate the radio properties with cluster
properties. 
Clusters with halos or relics are characterized by strong dynamical activity
related to merging processes. Moreover the most powerful radio halos and
relics are detected in clusters with the highest X-ray luminosity.
However, halos are not present in all merger clusters. To take into
account also upper limits to the radio emission for those clusters where a
radio halo is not detected, Feretti (2005) presented the correlation between
the average radio surface
brightness of the radio halo versus the cluster X-ray
luminosity. In Figure 4 we present this correlation including recent new
data.
It is interesting to note that upper limits are consistent with the correlation
suggesting that low X-ray luminosity clusters might host faint radio halos
which could be detected only by future telescopes as LOFAR, LWA, and SKA.

\begin{figure}
{\includegraphics[angle=0,width=9.0cm,viewport=0 25 592 440,clip]
{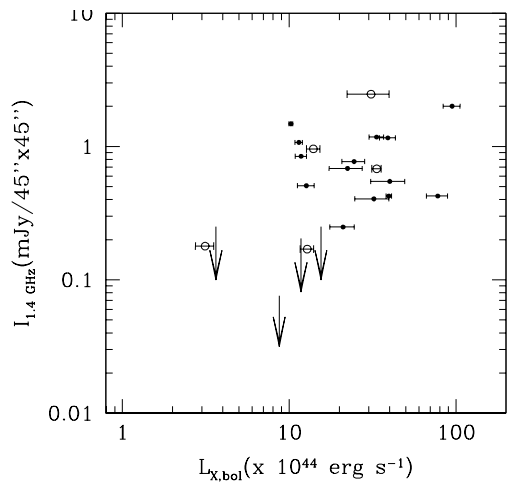}}
\caption{Halo radio brightness at 1.4 GHz versus cluster 
bolometric X-ray
luminosity.
Upper limits (arrows) are A119, A399, A2111, RXCJ1234.2+0947. Empty circles are
small size ($<$ 1Mpc) halos.}
\label{a1758}
\end{figure}

\end{document}